\documentclass[pre,twocolumn,twoside,showpacs,showkeys,floatfix]{revtex4}
\usepackage{dcolumn}%
\usepackage{bm}%
\usepackage{graphicx,textcomp}
\usepackage{amssymb,amsfonts,amsmath}
\usepackage{color}

\newcolumntype{P}[1]{>{\centering\arraybackslash}p{#1}}
\newcolumntype{M}[1]{>{\centering\arraybackslash}m{#1}}

\begin{document}

\title{Combinatorial solutions to generalized electrorheological kernel aggregation} 

\author{Micha\l{} \L{}epek\footnote{Corresponding author: lepek@if.pw.edu.pl}, Agata Fronczak, Piotr Fronczak}

\affiliation{Faculty of Physics, Warsaw University of Technology, Koszykowa 75, PL-00-662 Warsaw, Poland}

\date{\today}



\begin{abstract}
For this paper, we studied the time evolution of a system of coagulating particles under a generalized electrorheological (ER) kernel with real power, $K\left(i,j\right) = \left( \frac{1}{i}+\frac{1}{j} \right)^\alpha$, and monodisperse initial conditions. We used a combinatorial framework in which time and cluster sizes were discrete and the binary aggregation governed the time evolution of the system. We modified a previously-known solution for the constant kernel to cover the generalized ER kernel and used it in the framework to obtain the exact expression for the cluster size distribution (the average number of particles of a given size) and the standard deviation. Our theoretical solution is validated by a comparison to numerically simulated results for several values of  $\alpha$ and to the experimental data of coagulating polystyrene particles. Theoretical predictions were accurate for any time of the aggregation process and for a wide range of $\alpha$.
\end{abstract}

\maketitle

\section{Introduction}\label{SectIntro}

Coagulation processes (also known as aggregation or coalescence) are widespread in nature. They govern many everyday--life phenomena (such as blood coagulation, milk curdling, and cloud formation) and are of great interest in physics \cite{paper1, paper2, paper3}, chemistry \cite{paper4, paper5, paper6}, biology \cite{paper7}, and mathematics \cite{paper8,paper9,paper10}. Several technological applications are based on coagulation, including the formation of aerosols \cite{Drake_1972, Pruppacher_1978} and polymers \cite{Stockmayer_1943} and material processing \cite{Wattis_2004, Harris_2001}.

ER fluids are colloidal suspensions of electrically-active particles in an insulating fluid \cite{Winslow_1949}. If an external electric field is applied to the system, the particles acquire electric dipoles and irreversibly aggregate into linear chains oriented in the direction of the field, considerably changing the rheological and optical properties of the suspension. One important property to observe is the fluid's fast response time. Typical ER fluid can change consistency from a liquid to a gel in few milliseconds. Due to this advantage, ER fluids are sometimes called as ``smart'' materials and are used in several applications, such as hydraulic valves \cite{Simmonds_1991}, clutches \cite{Monkman_1997}, brakes \cite{Seed_1986}, shock absorbers \cite{Stanway_1996}, abrasive polishing \cite{Kim_2003}, and tactile displays \cite{Monkman_1992, Liu_2005}. Recently, the coagulation of colloidal particles in the presence of an external magnetic field has also been studied \cite{Bossis_2013, Reynolds_2016}.

The coagulation process can be regarded as the evolution of a closed system of clusters merging irreversibly as a result of binary collisions (coagulation acts) according to the general scheme 
\begin{equation} \label{EQ1}
   \left(i\right)+\left(j\right){{\stackrel{K\left(i,j\right)}{\longrightarrow}}}\left(i+j\right)
\end{equation}

\noindent where $\left(i\right)$ stands for a cluster of mass $i$ and $K\left(i,j\right)$ is the coagulation kernel representing the rate of the process. As the process is irreversible, the number of clusters decreases in time and eventually all of the clusters join into one single cluster.

The classic (deterministic) model for an aggregation process is the Smoluchowski aggregation equation \cite{paper12, paper13,paper14,paper15,paper16, paper17,paper18,paper19}. The advantage of this approach is that explicit analytical solutions are known for particular kernels (e.g., constant, multiplicative, or additive). However, this approach has several weaknesses. In particular, it requires the following assumptions: an infinite size for the system considered and continuous cluster concentrations. Because this does not account for small systems and because the number of clusters in large systems decreases significantly over time, this equation becomes unusable. These problems can be especially observed in the case of so-called ``gelling'' kernels. Moreover, the solutions arising from the Smoluchowski aggregation equation are stochastically incomplete and describe only the average behavior of clusters without providing any information on deviations.

For these reasons, a stochastic approach to studying finite coagulating systems has been proposed in recent literature \cite{paper20, paper21, paper22, paper23, paper24}. This idea for solving aggregating systems uses combinatorial equations to derive exact expressions for cluster size distribution in time. Thus far, this combinatorial approach was used to find solutions to constant, additive, and multiplicative kernels \cite{2018_PREFronczak, 2019_ROMP_Lepek, grassberger1, grassberger2}. The combinatorial framework proposed in \cite{2019_ROMP_Lepek} is of particular interest as it not only provides the expressions for the standard deviation of the mean values of cluster size distribution, but also can be extended to cover other arbitrary kernels if the aggregation rate $K$ can be written in the appropriate form. Additionally, it has been proven to be effective for systems with constant, multiplicative, and additive kernels and monodisperse initial conditions.

Several other works \cite{Miyazima_1987, Fraden_1989, Melle_2001, Mimouni_2007, Wattis_2009} studied the kinetics of irreversible aggregation in ER fluids. In the latest one \cite{Wattis_2009}, the authors analyzed a system of aggregating polystyrene particles. They showed that the process is governed by the coagulation kernel with negative powers of cluster sizes
\begin{equation} \label{electrorheological_kernel}
K\left(i,j\right) = \frac{1}{i}+\frac{1}{j}
\end{equation}

\noindent and used the reduction of the classic Smoluchowski aggregation equation to a similarity solution in a large-time limit. These theoretical results were compared to the experimental data and their limited accuracy was observed.

In this paper, we used the combinatorial approach proposed in \cite{2019_ROMP_Lepek} to solve the generalized form of the ER kernel (Eq.~(\ref{electrorheological_kernel})). The generalized form is obtained by using a real power:
\begin{equation} \label{generalized_electrorheological_kernel}
K\left(i,j\right) = \left( \frac{1}{i}+\frac{1}{j} \right)^\alpha.
\end{equation}

Many real processes can be approximated by using formulas with power factors. In \cite{Wattis_2009}, other power forms are proposed for further investigation. In the next sections, we solve the coagulation process Eq.~(\ref{EQ1}) using the kernel Eq.~(\ref{generalized_electrorheological_kernel}) with an arbitrary real power and compare the results with numerical simulations for several values of $\alpha$. We believe that these power-generalized form can be better used with experimental data.

The combinatorial framework we used requires the following assumptions: (i) monodisperse initial conditions, (ii) discrete time, and (iii) one coagulation act occurring in each time step. ER fluid is especially well-suited for this approach as its aggregation generally meets condition (i). Successive steps of the coagulation process define the space of available states, and the probability distribution over the state space is determined by studying the possible growth histories of clusters using combinatorial expressions. Then, the expressions for cluster size distribution and its standard deviation are derived. Although the combinatorial approach may seem complex, it is vital to emphasize that most of the equations are provided by the framework at once. The only issue that we worked on in this contribution involved transforming a recurrent expression for the number of possible internal states of a cluster to a non-recurrent form.

The paper is organized as follows. Section 2 presents the basics of the combinatorial approach. Section 3 provides a detailed description of our method for calculating the number of possible internal states of a cluster for the generalized ER kernel (Eq.~(\ref{generalized_electrorheological_kernel})). Section 4 compares the results of theoretical predictions to numerical simulations. Section 5 gives concluding remarks and describes possible extensions to this work.

\section{Combinatorial approach essentials} \label{Essentials}

Here, we briefly describe the  essentials of the combinatorial approach to coagulating systems \cite{2018_PREFronczak, 2019_ROMP_Lepek} that we used in this work. When investigating the aggregating system with this methodology, we assume discrete time and monodisperse initial conditions (all of the clusters are monomers of size of one). A single coagulation act occurs in one time step. Then, the total number of clusters, $k$, at time $t$ is
\begin{equation} \label{EQ2}
 k=N-t,  
\end{equation}

\noindent where $N$ is the total number of monomeric units in the system. As this number does not change during the evolution of the system, $N$ is equivalent to the initial number of clusters (preservation of mass). The state of the system at time $t$ is described by
\begin{equation} \label{EQ3}
 \mathrm{\Omega }\left(t\right)=\left\{n_1,n_2,\dots ,n_g,\dots ,n_N\right\},
\end{equation}

\noindent where $n_g\ge 0$ stands for the number of clusters of mass $g$ (therefore $g$ is the number of monomeric units included in the cluster) and $n_1$ corresponds to monomers, $n_2$ to dimers, $n_3$ to trimers, and so on. During the coagulation process the sequence $\left\{n_g\right\}$ is not arbitrary and satisfies following conditions corresponding to the preservation of the number of monomeric units in the system:
\begin{equation} \label{constraints}
 \sum^N_{g=1}{n_g=k} \;\;\;\; \textrm{and} \;\;\;\; \sum^N_{g=1}{{g\ n}_g=N}.
\end{equation}

According to \cite{2018_PREFronczak, 2019_ROMP_Lepek}, there are three origins of combinatorial expressions to model the aggregation process. The first one results from the fact that the set of monomers can be divided into subsets in a specific number of ways. The second origin results from distributing coagulation acts of the process in different time steps. The third aspect of combinatorial description covers the number of ways in which a given cluster could be created (the number of possible histories of a cluster). It has been shown (in \cite{2018_PREFronczak}) that by combining these expressions together one can derive the average number of clusters $\left\langle n_s\right\rangle$ of a given size $s$ as
\begin{equation} \label{ns_general}  
 \left\langle n_s\right\rangle =\binom{N}{s}{\omega }_s\frac{B_{N-s,k-1}\left(\left\{{\omega }_g\right\}\right)}{B_{N,k}\left(\left\{{\omega }_g\right\}\right)}
\end{equation}

\noindent or, for simplicity's sake, as
\begin{equation} \label{omega_s}  
\omega_s = \frac{x_s}{(s-1)!} \;\;\;\; \textrm{and} \;\;\;\; \left\{\omega_g\right\} =  \left\{ \frac{x_g}{(g-1)!} \right\} .
\end{equation}

Here, we need to better explain Eq.~(\ref{ns_general}). It describes the average number of clusters of size $s$ after $t$ steps of the aggregation process. Although $t$ is not explicitly present in the equation, $k$ plays $t$'s role, as $k$ is the total number of clusters in the system and decreases linearly with time.

We must also explain the difference between $\omega_s$ and $\omega_g$. The first one is a single value and depends on cluster size $s$, while $\{\omega_g\}$ is a sequence not dependent on $s$, where $g$ varies from $1$ to $N-k+1$ (i.e., to $t+1$).



The sequence $\left\{\omega_g\right\}$ is used to calculate so--called partial (or incomplete) Bell polynomials. They are defined as
\begin{multline} \label{Bell_polynom_def}
 B_{N,k}\left(z_1,z_2,\dots ,z_{N-k+1}\right)=B_{N,k}\left(\left\{z_g\right\}\right) \\ =N!\sum_{\left\{n_g\right\}}{\prod^{N-k+1}_{g=1}{\frac{1}{n_g!}{\left(\frac{z_g}{g!}\right)}^{n_g}}} 
\end{multline} 

\noindent where the summation is taken over all non-negative integers $\left\{n_g\right\}$ that satisfy Eq.~(\ref{constraints}). Bell polynomials are a useful tool in combinatorics as they provide detailed information about the partition of an arbitrary set. Several computational environments implement Bell polynomials (e.g. Mathematica). In the Appendix \cite{appendix}, we provide efficient equation to calculate partial Bell polynomials used in this research.

As mentioned in the Introduction, not only can the average number of clusters be estimated in the combinatorial framework, but so can the corresponding standard deviation of this average,
\begin{equation} \label{std_dev_general}  
 {\sigma }_s=\sqrt{\left\langle n_s\left(n_s-1\right)\right\rangle +\left\langle n_s\right\rangle -{\left\langle n_s\right\rangle }^2}
\end{equation}

\noindent where, for $2s \leqslant N$,
\begin{equation} \label{std_dev_general_addition}  
 \left\langle n_s\left(n_s-1\right)\right\rangle =\binom{N}{s,s}{{\omega }_s}^2\frac{B_{N-2s,k-2}\left(\left\{{\omega }_g\right\}\right)}{B_{N,k}\left(\left\{{\omega }_g\right\}\right)}\
\end{equation}

\noindent with $\binom{N}{s,s}=\binom{N}{s}\binom{N-s}{s}$ and $\left\langle n_s\left(n_s-1\right)\right\rangle = 0$ for other cases.

Up to this point, the combinatorial equations were generic and applicable to any kind of aggregation kernel. Therefore, since this approach focuses on the reaction kernel $K$, the kernel is only used to calculate the number of possible histories of a cluster of given size $x_g$ (it can be also regarded as a number of possible internal states of a cluster). This number is unambiguously defined by the kernel.

For the most basic case of the constant kernel, the number $x_g$ is specified by the recurrent expression \cite{2019_ROMP_Lepek}
\begin{equation} \label{xg_recurrent_definition}
 x_g=\frac{1}{2}\sum^{g-1}_{k=1}\binom{g}{k}\binom{g-2}{k-1}x_kx_{g-k}
\end{equation}

\noindent where $x_k$ and $x_{g-k}$ are 
the numbers of ways to create the two merging clusters. This recurrent expression is build as follows. The first Newton symbol, $\binom{g}{k}$, denotes the number of ways of choosing clusters of size $k$ out of $g$ monomers as we can divide the cluster of size $g$ into subclusters of size $k$ and size $(g-k)$ in exactly $\binom{g}{k}$ ways. The second Newton symbol, $\binom{g-2}{k-1}$, stands for the fact that the coagulation acts of clusters of sizes $(g-k)$ and $k$ could appear in different time steps. Coagulation acts related to the creation of a cluster of size $k$ could occur in $k-1$ time steps out of the total number of $g-2$ time steps needed to create clusters of sizes $k$ and $(g-k)$. The sum is taken over the possible pairs of clusters that can result in the merged cluster of size $g$. The factor of $\frac{1}{2}$ is used to prevent double counting of coagulation acts.

In the next section, we will modify Eq.~(\ref{xg_recurrent_definition}) to describe the generalized ER kernel and transform it to the non-recurrent form that can be used in Eq.~(\ref{ns_general}) to calculate $\left\langle n_s\right\rangle$.

\section{Calculating $x_g$ for generalized electrorheological kernel}

At this point, we must modify the recurrent expression for $x_g$ to cover the ER kernel. In the previous work \cite{2019_ROMP_Lepek}, we have shown that $x_g$ for the constant kernel can be modified to cover the additive and multiplicative kernels by multiplying the right-hand side of Eq.~(\ref{xg_recurrent_definition}) by a specific factor. This factor is the kernel reaction rate $K$ itself, but translated into the variables $g$ and $k$ used in Eq.~(\ref{xg_recurrent_definition}). This multiplying is the result of the fact that the probability of the coagulation act is proportional to $K$. Therefore, the general expression for $x_g$ for any kernel is
\begin{equation} \label{xg_recurrent_definition_general}
 x_g=\frac{1}{2}\sum^{g-1}_{k=1}\binom{g}{k}\binom{g-2}{k-1}x_kx_{g-k}K(g,k).
\end{equation}

Please note that this multiplying is also consistent with the constant kernel, since $K=1$ for the constant kernel.

Now, we must translate the ER kernel, $K\left(i,j\right) = \left( \frac{1}{i}+\frac{1}{j} \right)^\alpha$, to the language of $g$ and $k$. The ER kernel can be rewritten to
\begin{equation} \label{Kij_rewritten}
 K\left(i,j\right) = \left( \frac{i+j}{ij} \right)^\alpha.
\end{equation}

Bearing in mind that $i$ and $j$ are the masses of two merging clusters, $g$ is the mass of the resulting cluster, and $k$ is the mass of one of the subclusters (e.g., $i=k$), we can write
\begin{equation} \label{Kgk}
 K\left(g,k\right) = \left( \frac{k+(g-k)}{k(g-k)} \right)^\alpha = 
 \left( \frac{g}{k(g-k)} \right)^\alpha.
\end{equation}

Thus, the recurrent expression for $x_g$ for the ER kernel is
\begin{equation} \label{xg_recurrent_electro}
 x_g=\frac{1}{2}\sum^{g-1}_{k=1}\binom{g}{k}\binom{g-2}{k-1}x_kx_{g-k} \left( \frac{g}{k(g-k)} \right)^\alpha.
\end{equation}

Expanding the Newton symbols and substituting $y_g=\frac{x_g}{g!(g-1)!g^\alpha}$ we obtain
\begin{equation} \label{electro_2} 
 (g-1)y_g=\frac{1}{2}\sum^{g-1}_{k=1}{y_ky_{g-k}}.
\end{equation}

Note that Eq.~(\ref{electro_2}) is exactly the same as for the constant kernel in \cite{2019_ROMP_Lepek} with the only difference being the definition of $y_g$.This equation can be transformed to the explicit expression for $y_g$ using the generating function method. As it was solved in the previous work \cite{2019_ROMP_Lepek}, here we will use the known solution,
\begin{equation} \label{yg_general_solution}  
 y_g=\frac{1}{2^{g-1}},
\end{equation}

\noindent which, with the substitution, results in this equation:
\begin{equation} \label{electro_3}  
 \frac{1}{2^{g-1}} = \frac{x_g}{g!g^\alpha\left(g-1\right)!}.
\end{equation}

Finally, the non-recurrent form of $x_g$ for the generalized ER kernel is
\begin{equation} \label{electro_xg}  
 x_g=\frac{g!g!g^{\alpha-1}}{2^{g-1}}.
\end{equation}

The ``regular'' ER process is described by $x_g$ with $\alpha=1$. It is worth noting that for $\alpha=0$, the solution fully corresponds to the solution for the constant kernel, known from \cite{2019_ROMP_Lepek}.

\begin{figure*}[ht] \label{Figure_1}
\includegraphics[width=1\textwidth]{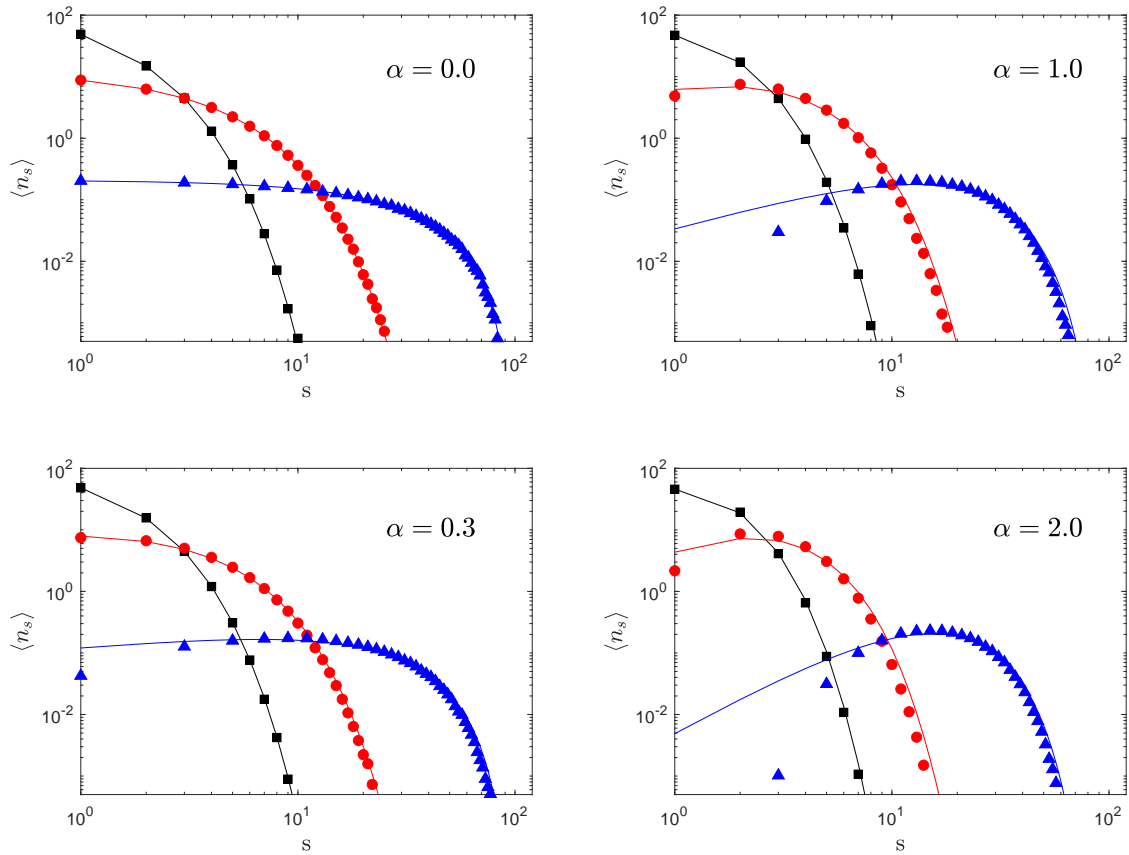}
\caption{Numerical simulations vs. theoretical calculations of the average number of clusters of the size $s$, $\langle n_s \rangle$, for the generalized electrorheological kernel with different values of $\alpha$ ($\alpha=0.0$, $\alpha=0.3$, $\alpha=1.0$, $\alpha=2.0$). For $\alpha=0.0$, the results fully correspond to the additive kernel. For $\alpha=1.0$, we obtain the results for the "regular" form of the electrorheological kernel. Solid lines represent theoretical predictions based on combinatorial equations. Circles, squares and triangles correspond to the results obtained by numerical simulation. The number of monomers in the system is $N=100$. Three stages of the aggregation process are presented: $t=30$ (squares, black), $t=70$ (circles, red) and $t=95$ (very late stage of the process, triangles, blue). For each case, ${10}^5$ independent simulations were performed.}
\end{figure*}

\section{The final expression for $\left\langle n_s\right\rangle$ compared to numerical results and experimental data}

Theoretical solutions arising from our combinatorial equations have been compared to the results obtained by numerical simulations. These solutions were obtained using the general expression for $\left\langle n_s\right\rangle$, Eq.~(\ref{ns_general}), and the expression for $x_g$, which was the number of possible internal states of a cluster of size $g$ (Eq.~(\ref{electro_xg})) derived in the previous section. The expression for $x_g$ is used in Eq.~(\ref{ns_general}) twice: once as $x_s$,
\begin{equation} \label{omega_s_later}  
\omega_s = \frac{ s!s!s^{\alpha-1} }{ 2^{s-1}(s-1)! } = 
  \frac{ s!s^{\alpha} }{ 2^{s-1} },
\end{equation}

\noindent and once in the sequence $\left\{\omega_g\right\}$,
\begin{equation} \label{omega_g_later}  
 \left\{\omega_g\right\} =  \left\{ \frac{ g!g!g^{\alpha-1} }{  2^{g-1}(g-1)!  } \right\}  = \left\{ \frac{ g!g^{\alpha} }{ 2^{g-1} } \right\},
\end{equation}

\noindent where $g$ changes from $1$ to $N-k+1$ (i.e., to $t+1$).

Therefore, the final expression for the average number of clusters of a given size, $\left\langle n_s\right\rangle$, as it is plotted in the figures, takes the form
\begin{equation} \label{ns_solved_final}  
 \left\langle n_s\right\rangle =\binom{N}{s} \frac{ s!s^{\alpha} }{ 2^{s-1} } \frac{ B_{N-s,k-1}\left(  \left\{ \frac{ g!g^{\alpha} }{ 2^{g-1} } \right\}   \right) }{ B_{N,k}\left(\left\{  \frac{ g!g^{\alpha} }{ 2^{g-1} }  \right\}\right)}
\end{equation}

\noindent where, again, $g$ changes from $1$ to $N-k+1$ (i.e., to $t+1$).

Although Eq.~(\ref{ns_solved_final}) is sufficient and fully defines $\left\langle n_s\right\rangle$, for this particular $\left\{\omega_g\right\}$, it can be further simplified. Using the relation \cite{Comtet_1974}
\begin{multline} \label{Bell_relation_1}  
B_{N,k}\left(abz_1,ab^2z_2,\dots,ab^{N-k+1}z_{N-k+1}\right) \\ = 
 a^kb^NB_{N,k}\left(z_1,z_2,\dots,z_{N-k+1}\right),
\end{multline}

\noindent we obtain simplified form of Eq.~(\ref{ns_solved_final}),
\begin{equation} \label{ns_solved_final_simplified}  
 \left\langle n_s\right\rangle =\binom{N}{s}  s!s^{\alpha}  \frac{ B_{N-s,k-1}\left(  \left\{  g!g^{\alpha} \right\}   \right) }{ B_{N,k}\left(\left\{  g!g^{\alpha}   \right\}\right)}.
\end{equation}

Moreover, to avoid expensive calculations of Bell polynomials, further simplification can be done for the ''regular'' ER kernel ($\alpha=1$) using the identity relation with the so--called falling factorial \cite{Wang_2009}. This relation is
\begin{multline} \label{Bell_relation_2}  
 B_{N,k}\left(\left\{  g!g   \right\}\right)  = 
  \frac{1}{k!} \sum^{k}_{j=0} (-1)^{k-j} \binom{k}{j} (j+k+N-1)_N
\end{multline}

\noindent and the falling factorial is defined as
\begin{equation} \label{falling_factorial} 
(a)_b = a(a-1) \dots (a-b+1).
\end{equation}

Using the relation (\ref{Bell_relation_2}), we can transform $\left\langle n_s\right\rangle$ for the ''regular'' ER kernel to the form which does not consist of Bell polynomials,
\begin{multline} \label{ns_solved_final_simplified_alpha_1}  
 \left\langle n_s(\alpha=1)  \right\rangle \\  =
 \binom{N}{s}  s!sk  \frac{  \sum^{k-1}_{j=0} (-1)^{k-1-j} \binom{k-1}{j} (j+k+N-2)_N  }{  \sum^{k}_{j=0} (-1)^{k-j} \binom{k}{j} (j+k+N-1)_N  }
 .
\end{multline}

In Figure 1, the results for several values of power $\alpha=0.0, 0.3, 1.0$, and $2.0$ are presented. There are three data series contained in each plot, corresponding to three different stages of the system evolution. The initial number of monomers is $N=100$. The first series ($t=30$) corresponds to the early stage of the aggregation process; the second series ($t=70$) covers the later stage of the process; and the third series shows the average numbers of clusters near the end of the process ($t=95$), which is five steps before the moment when all of the particles are joined into one single cluster. For low values of $\alpha$ ($0.0$ and $0.3$), the theoretical prediction reflects the simulation with excellent precision, even for the latest phase of the process ($t=95$). A somewhat precise result can also be obtained also for $\alpha=1.0$ (``regular'' ER kernel) and $\alpha=2.0$, although, the predicted average values of $\langle n_s \rangle$ for the smallest clusters (i.e. $s<5$) are higher than the values calculated from the simulation. Notably, the highest point of the cluster size distribution is always modeled precisely by the theoretical curves (in this regime, it can be regarded as the exact solution).

It was proposed in \cite{2019_ROMP_Lepek} that the disagreement between theoretical and numerical results for the later phases for the gelling kernels is caused by the appearance of a giant gel cluster in the system. When the system crosses the gelling point, this giant cluster changes the probabilities in the system, which is not covered in the combinatorial expressions. Although the ER kernel is not obviously gelling (like, e.g., multiplicative kernel), the largest clusters are the clusters of moderate $s$ (especially for higher $\alpha$). Thus, there is a relatively high number of large clusters in the system. For the gelling kernels solved in this combinatorial approach, the theoretical solutions (for the latest stages of the process) are then ``delayed'' in comparison to the numerical simulation. This effect can be seen (slightly) for $\alpha=2.0$ in the case of the ER kernel.

In Figure 2, we present standard deviation estimates given by the combinatorial approach, Eq.~(\ref{std_dev_general}), and compare them to the standard deviation calculated for the data obtained by simulation. The results for three values of $\alpha$ are presented. It can be seen that the combinatorial estimates behave similarly to the average number of clusters, being precise for the lowest $\alpha$ and somewhat precise for the ``regular'' ER kernel. For the highest $\alpha=2.0$ the deviation estimates are only approximate for the lowest $s$ but still acceptable for the rest of the distribution. Minimal variations from the numerical data also occur for the highest $s$ (see Figure 2b and 2c).

Additionally, to compare our theoretical predictions to the experimental data, we have adopted the data gained by Wattis and Mimouni \cite{Wattis_2009}. These authors performed an experiment in which polystyrene particles were suspended in the liquid (water and heavy water). When the alternating electric field was applied, the particles started to coagulate into chains, a process was observed by the microscope and the camera. In their work, the authors provided the raw data (the probabilities of finding clusters of a given size) for four time steps of the process ($t1=1 min$, $t2=3 min$, $t3=5 min$, and $t4=7 min$). The total number of monomer units in the system was estimated by the authors as 173, 141, 118, and 121 (for several time steps). Thus, we took the average of these numbers, $N=138$. We also normalized the raw data probabilities in such a way, that the number of monomers $N$ was preserved for each time step. Then, having the number of clusters in the system for several time steps, the time in minutes could be translated into the time counted as binary coagulation acts. In this way, we obtained: $t1=83$, $t2=95$, $t3=107$, and $t4=117$. As we had $N$ and $t$, we could instantaneously plot the theoretical curves versus the data. Please see Figure 3 for the comparison of the data and the theoretical predictions for the ``regular'' ER kernel ($\alpha=1$). Mostly, the data points for all of the time points shown stay inside the area limited by the theoretically predicted standard deviation. Evidently, the combinatorial predictions described this coagulation process properly.

The methods and issues related to the numerical studies and calculations are outlined in the Appendix \cite{appendix}. The code used for the simulations and for the theoretical predictions is available at https://github.com/mlepek/aggregation.

\section{Concluding remarks}\label{SecSum}

As previously know, an exact combinatorial approach with recursive equations gives excellent results for simple kernels as the constant or additive \cite{2019_ROMP_Lepek}. In this research, we used the combinatorial approach for determining the number $x_g$ in all possible histories of a cluster of a given size to cover a generalized ER kernel with real power. This approach can be useful in studies on real systems as they are often are likely to behave as power functions. We showed the performance of these combinatorial solutions by comparing the results of numerical simulations with a system size of $N=100$. The performance varied for different values of $\alpha$ and different times of coagulation. In the early stages of the process ($t=30$), we obtained excellent results for all of kernel forms considered. In the later stages ($t=70$), theoretical predictions followed the numerical simulations with excellent ($\alpha=0.0$, $\alpha=0.3$, $\alpha=1.0$) or, at least, high precision ($\alpha=2.0$). In the last stages of the process, just before merging into one single cluster ($t=95$), the results remained very good for all of the cases, preserving the top peak of the curve with excellent precision. However, in the case of $\alpha=2.0$ and the regime of the smallest clusters ($s=1,2,3,4$), theoretical predictions were higher than the numerical results. Similarly, the precision of standard deviation estimates decreased for low $s$ and higher $\alpha$.

\begin{figure}[ht!] \label{Figure_2}
\includegraphics[scale=0.88]{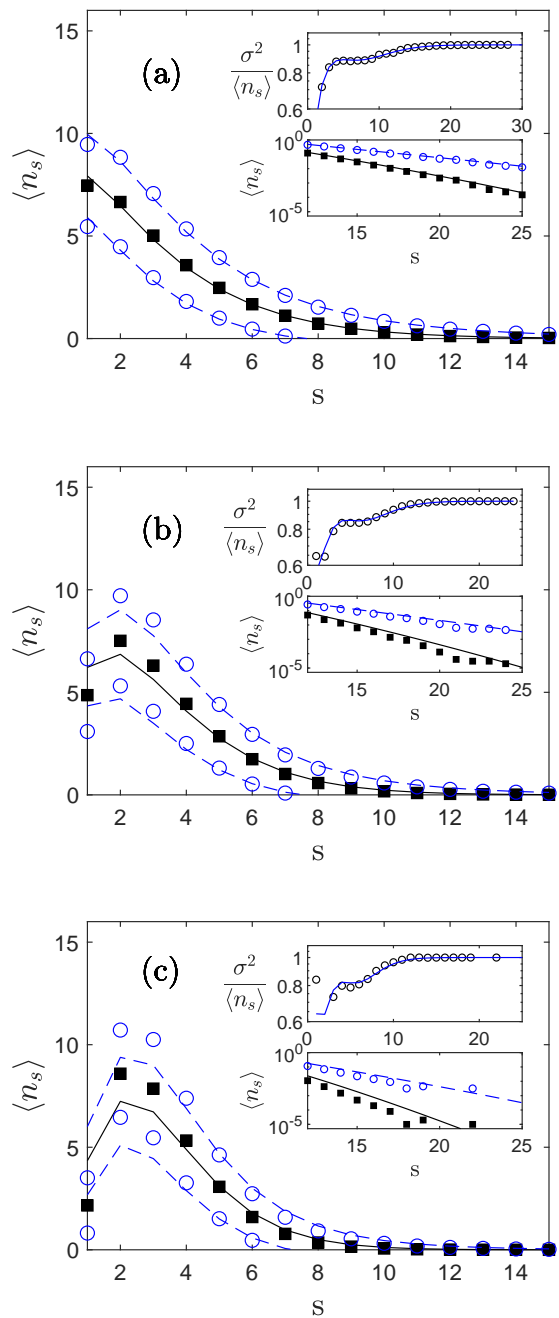}
\caption{Standard deviation predicted by the theoretical model, Eq.~(\ref{std_dev_general}), vs. standard deviation obtained by the simulation for the electrorheological kernel with $t=70$ and several values of $\alpha$: (a) $\alpha=0.3$, (b) $\alpha=1.0$, and (c) $\alpha=2.0$. Solid and dashed lines represent combinatorial results for the average number of clusters of a given size, $\langle n_s\rangle$, and for standard deviation, respectively; squares and circles represent, respectively, the average number of clusters and standard deviation obtained by simulation. Upper inset figures: plots of variance divided by mean, $ \sigma^2/ \langle n_s\rangle$ (solid lines for theorhetical prediction and circles for numerical data). Lower inset figures: plots of $\langle n_s\rangle$ for higher $s$ (logarithmic scale). For each case, ${10}^5$ independent simulations were performed.}
\end{figure}

\begin{figure*}[ht] \label{Figure_3}
\includegraphics[width=1\textwidth]{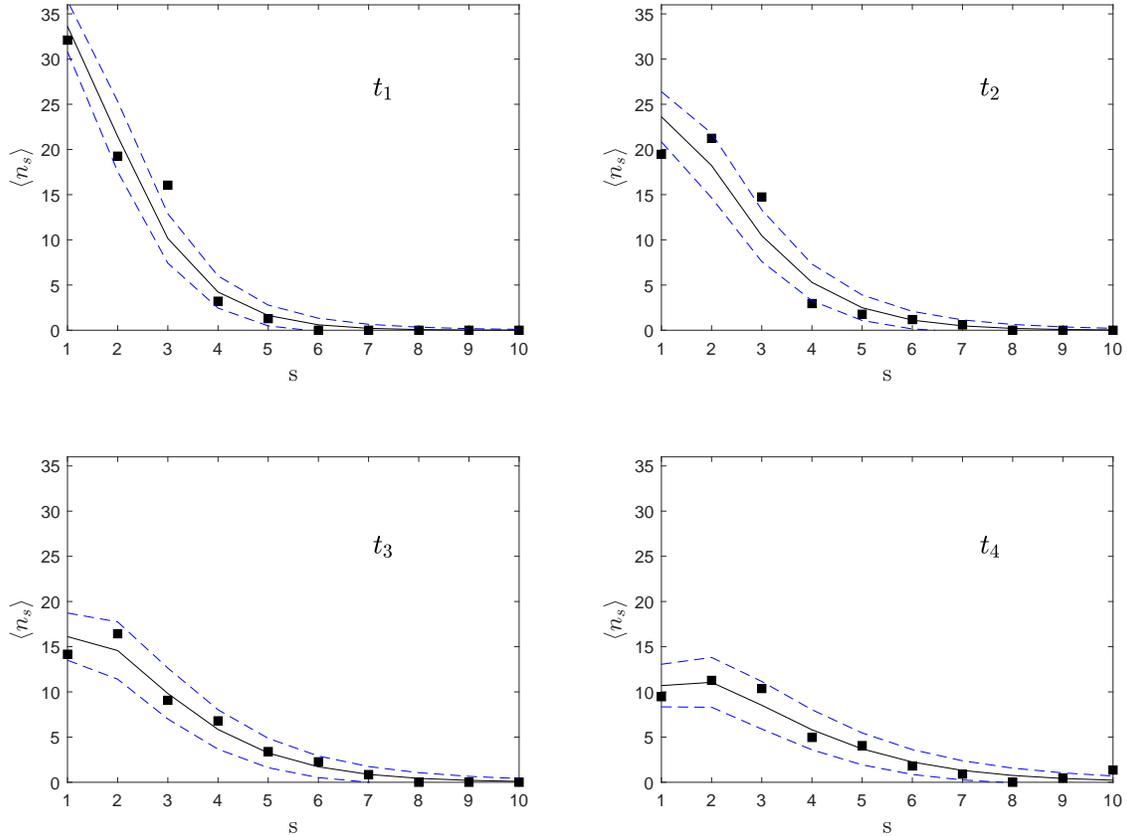}
\caption{Theoretical calculations of the average number of clusters of size $s$, $\langle n_s \rangle$, for the electrorheological kernel ($\alpha=1$) compared to the experimental data of coagulating polystyrene particles taken from \cite{Wattis_2009}. Four time--steps of the process are presented: $t_1=1 min$ ($t=83$ in the time counted as binary aggregation acts), $t_2=3 min$ ($t=95$), $t_3=5 min$ ($t=107$), and $t_4=7 min$ ($t=117$). The number of monomers in the system was estimated as $N=138$. Solid and dashed lines represent theoretical predictions of $\langle n_s \rangle$ and its standard deviation, respectively. Squares represent experimental data points.}
\end{figure*}

In previous work on ER aggregation \cite{Wattis_2009}, Wattis and Mimouni considered the ``regular'' form of the kernel (\ref{generalized_electrorheological_kernel}) with $\alpha=1$ and, deriving from the Somuchowski's equation, they obtained theoretical results that compare with the experimental data. This data has been obtained in the setup \cite{Mimouni_2007} that used a small transparent cell containing a colloidal suspension of spherical polystyrene particles in a mixture of water (H\textsubscript{2}O) and heavy water (D\textsubscript{2}O). Wattis and Minouni showed that the data shows good agreement with their theory for the later stages of the aggregation process. However, there was a considerable difference between the theory and the data at the initial time steps (please cf. Fig. 4 in \cite{Wattis_2009}). In Fig. 3 we show that using the combinatorial approach presented herein we can model the process with high precision for any time. Particularly, the quality of the solution Eq.~(\ref{ns_solved_final_simplified_alpha_1}) is demonstrated in comparison with the experimental data.

The results presented in this work prove the evident generality of the combinatorial approach. This generality is a new quality in aggregation studies, since the only element needed to cover another type of kernel is the number of possible histories of a cluster, $x_g$, which can sometimes be easily obtained. All of the theoretical solutions are also supplied by the information on the standard deviation for each kernel, providing ``stochastic completeness.''

What can we suggest to be done further? For sure, an obvious task is developing the approach to cover initial conditions other than monodisperse conditions as it was done for the product kernel in the Marcus-Lushnikov framework \cite{2019_PRE_Fronczak}. It would be of particular interest in view of the known sensitivity of the coagulation process to its initial conditions \cite{2004_Menon}. Of course, there are also other interesting questions. Why the methodology used here gives exact results for some kernels, while approximate for others? Is it possible to modify the combinatorial expressions to obtain better results for the time after the phase transition in case of the gelling kernels? Answering these questions would have a great impact on the coagulation theory as it would allow to model large variety of systems in arbitrary conditions with high precision.

\acknowledgments
This work has been supported by the National Science Centre of Poland (Narodowe Centrum Nauki) under grant no.~2015/18/E/ST2/00560 (A.F. and M.\L.).


\begin{thebibliography}{99}

\bibitem{paper1}
   P.L.~Krapivsky, S.~Redner and E.~Ben-Naim: {\em A Kinetic View of Statistical Physics\/} (Chapter 5), Cambridge University Press,
   New York 2010.

\bibitem{paper2}
J.A.D.~Wattis: {\em Physica D\/} {\bf 222}, 1 (2006), An introduction to mathematical models of coagulation--fragmentation processes: A discrete deterministic mean--field approach.

\bibitem{paper3}
F.~Leyvraz: {\em Phys. Rep.\/} {\bf 383}, 95 (2003), Scaling theory and exactly solved models in the kinetics of irreversible aggregation.

\bibitem{paper4}
   H.~Sontag and K.~Strenge: {\em Coagulation Kinetics and Structure Formation\/}, Plenum Press,
   New York 1987.

\bibitem{paper5}
   F.~Family and D.P.~Landau: {\em Kinetics of Aggregation and Gelation\/}, North--Holland,
   Amsterdam 1984.

\bibitem{paper6}
   R.L.~Drake in: G.M. Hidy and J.R.~Brock (eds.): {\em Topics in Current Aerosol Researches\/} (Part II), Pergamon,
   New York 1972.

\bibitem{paper7}
   J.~Hein, M.H.~Schierup and C. Wiuf:
   {\em Gene Genealogies, Variation and Evolution -- A Primer in Coalescent Theory\/}, Oxford University Press,
   New York 2005.

\bibitem{paper8}
   J.~Bertoin:
   {\em Random Fragmentation and Coagulation Processes\/}, Cambridge University Press,
   Cambridge 2006.

\bibitem{paper9}
   J.~Pitman:
   {\em Combinatorial Stochastic Processes\/}, Springer--Verlag,
   Berlin 2006.

\bibitem{paper10}
D.J.~Aldous: {\em Bernoulli\/} {\bf 5}, 3 (1999), Deterministic and stochastic models for coalescence (aggregation and coagulation): a review of the mean field theory for probabilists.

\bibitem{Drake_1972}
   R.L.~Drake:
    {\em A general mathematical survey of the coagulation equation, in: International Reviews in Aerosol Physics and Chemistry\/}, Pergamon Press,
   New York 1972.
   
\bibitem{Pruppacher_1978}
   H.R.~Pruppacher, J.D.~Klett:
    {\em Microphysics of clouds and precipitation\/}, Reidel,
   Dodrecht 1978.

\bibitem{Stockmayer_1943}
W.H.~Stockmayer: {\em J. Phys. Chem.\/} {\bf 11}, 45-55 (1943), Theory of molecular size distribution and gel formation in brached-chain polymers.

\bibitem{Wattis_2004}
J.A.D.~Wattis, D.G.~McCartney, T.~Gudmundsson: {\em J. Eng. Math.\/} {\bf 49}, 113-131 (2004), Coagulation equations with mass loss.
	
\bibitem{Harris_2001}
J.R.~Harris, J.V.~Wood, J.A.D.~Wattis: {\em Acta. Met.\/} {\bf 49}, 3991-4003 (2001), A comparison of potential models for mechanical alloying.
	
\bibitem{Winslow_1949}
W.M.~Winslow: {\em J. Appl. Phys.\/} {\bf 20}, 1137-1140 (1949), Induced fibration of suspensions.


\bibitem{Simmonds_1991}
A.J.~Simmonds: {\em IEE Proceedings D\/} {\bf 138}, 400-404 (1991), Electro-rheological valves in a hydraulic circuit.

\bibitem{Monkman_1997}
G.J.~Monkman: {\em Mechatronics\/} {\bf 7} (1) 27–36 (1997), Exploitation of compressive stress in electrorheological coupling.


\bibitem{Seed_1986}
M.~Seed, G.S.~Hobson, R.C.~Tozer, A.J.~Simmonds: {\em Proc. IASTED Int. Symp. Measurement, Sig. Proc. and Control.\/} Paper No. 105–092–1 (1986), Voltage-controlled Electrorheological brake.
	
\bibitem{Stanway_1996}
R.~Stanway, J.L.~Sproston, A.K.~El-Wahed: {\em Smart Mater. Struct.\/} {\bf 5}, 464–482 (1996), Applications of electro-rheological fluids in vibration control: a survey.	

\bibitem{Kim_2003}
W.B.~Kim, S.J.~Lee, Y.J.~Kim, E.S.~Lee: {\em International Journal of Machine Tools and Manufacture\/} {\bf 43} (1) 81-88 (2003), The electromechanical principle of electrorheological fluid-assisted polishing.
	
\bibitem{Liu_2005}
Y.~Liu, R.~Davidson, P.~Taylor: {\em Proceedings of SPIE. Smart Structures and Materials 2005: Smart Structures and Integrated Systems\/} {\bf 5764}, 92–99 (2005), Investigation of the touch sensitivity of ER fluid based tactile display.	


\bibitem{Monkman_1992}
G.J.~Monkman: {\em Presence: Teleoperators and Virtual Environments\/} {\bf 1} (2) 219–228 (1992), An Electrorheological Tactile Display.

\bibitem{Bossis_2013}
G.~Bossis, P.~Lancon, A.~Maunier, et al.: {\em Physica A: Statistical Mechanics and its Applications\/} {\bf 392} (7) 1567-1576 (2013), Presence: Kinetics of internal structures growth in magnetic suspensions.

\bibitem{Reynolds_2016}
   C.~Reynolds:: {\em Field induced assembly of paramagnetic colloidal particles\/}, PhD Thesis, University of Oxford, 2016.


\bibitem{paper12}
M.~Smoluchowski:
{\em Phys. Z.\/}
{\bf 17}, 557 (1916),
Drei vortrage uber diffusion bewegung und koagulation von kolloidteilchen.


\bibitem{paper13}
W.H.~White:
{\em Proc. Amer. Math. Soc.\/}
{\bf 80}, 273 (1980),
A global existence theorem for Smoluchowski's coagulation equation.


\bibitem{paper14}
R.M.~Ziff and G.~Stell:
{\em J. Chem. Phys.\/}
{\bf 73}, 3492 (1980),
Kinetics of polymer gelation.


\bibitem{paper15}
E.M.~Hendriks, M.H.~Ernst and R.M.~Ziff:
{\em J. Stat. Phys.\/}
{\bf 31}, 519 (1983),
Coagulation equation with gelation.


\bibitem{paper16}
P.G.J.~van~Dongen and M.H.~Ernst:
{\em J. Stat. Phys.\/}
{\bf 44}, 785 (1986),
On the occurrence of a gelation transition in Smoluchowski's coagulation equation.


\bibitem{paper17}
M.~Kreer and O.~Penrose:
{\em J. Stat. Phys.\/}
{\bf 75}, 389 (1994),
Proof of dynamical scaling in Smoluchowski's coagulation equation with constant kernel.


\bibitem{paper18}
F.~Leyvraz:
{\em Physica D\/}
{\bf 222}, 21 (2006),
Scaling theory for gelling systems: Work in progress.


\bibitem{paper19}
J.~Burnett and I.J.~Ford:
{\em J. Chem. Phys.\/}
{\bf 142}, 194112 (2015),
Coagulation kinetics beyond mean field theory using an optimized Poisson representation.


\bibitem{paper20}
A.H.~Marcus:
{\em Technometrics\/}
{\bf 10}, 133 (1968),
Stochastic coallescence.


\bibitem{paper21}
M.H.~Bayewitz, J.~Yerushalmi, S.~Katz and R.~Shinnar:
{\em J. Atmos. Sci.\/}
{\bf 31}, 1604 (1974),
The extent of correlations in a stochastic coalescence process.


\bibitem{paper22}
A.A.~Lushnikov:
{\em J. Colloid Interface Sci.\/}
{\bf 65}, 276 (1978),
Coagulation in finite systems.


\bibitem{paper23}
E.M.~Hendriks, J.L.~Spouge, M.~Eibl and M.~Schreckenberg:
{\em Z. Phys. B\/}
{\bf 58}, 219 (1985),
Exact solutions for random coagulation processes.


\bibitem{paper24}
A.A.~Lushnikov:
{\em Physica D\/}
{\bf 222}, 37 (2006),
Gelation in coagulating systems.


\bibitem{2018_PREFronczak} A.~Fronczak, A.~Chmiel, P.~Fronczak:
{\em Phys. Rev. E\/}
{\bf 97}, 022126 (2018),
Exact combinatorial approach to finite coagulating systems.
	
    

    

\bibitem{2019_ROMP_Lepek} A.~Fronczak, M.~\L{}epek, P.~Kukli\'nski, P.~Fronczak:
{\em Rep. Math. Phys.\/}
{\bf 84} (1) 117-130 (2019),
Exact combinatorial approach to finite coagulating systems through recursive equations.


\bibitem{grassberger1}
S.-W.~Son, C.~Christensen, G.~Bizhani, P.~Grassberger, and M.~Paczuski:
{\em Europhys. Lett.\/}
{\bf 95}, 58007 (2011),
Irreversible aggregation and network renormalization.

\bibitem{grassberger2}
S.-W.~Son, C.~Christensen, G.~Bizhani, P.~Grassberger, and M.~Paczuski:
{\em Phys. Rev. E\/}
{\bf 84}, 040102 (2011),
Exact solutions for mass-dependent irreversible aggregations.

\bibitem{Miyazima_1987}
S.~Miyazima, P.~Meakin, F.~Family:
{\em Phys. Rev. A\/}
{\bf 36} (3) 1421-1427 (1987),
Aggregation of oriented anisotropic particles.

\bibitem{Fraden_1989}
S.~Fraden, A.J.~Hurd, R.B.~Meyer:
{\em Phys. Rev. Lett.\/}
{\bf 63}, 2373-2376 (1989),
Electric-field-induced association of colloidal particles.

\bibitem{Melle_2001}
S.~Melle, M.A.~Rubio, G.G.~Fuller:
{\em Phys. Rev. Lett.\/}
{\bf 87}, 115501 (2001),
Time scaling regimes in aggregation of magnetic dipolar particles.

\bibitem{Mimouni_2007}
Z.~Mimouni:
{\em C. R. Physique \/}
{\bf 8}, 115-120 (2007),
Cinetique d'agregation en chaines dans une suspension colloidale soumise a un champ electrique alternatif.

\bibitem{Wattis_2009}
Z.~Mimouni, J.A.D.~Wattis:
{\em Physica A \/}
{\bf 388}, 1067-1073 (2009),
Similarity solution of coagulation equation with an inverse kernel.


\bibitem{Comtet_1974}
L.~Comtet:
{\em Advanced Combinatorics: The Art of Finite and Infinite Expansions\/}, Reidel Publishing Company, Dordrecht, Holland / Boston, U.S., 1974.

\bibitem{Wang_2009}
W.~Wang, T.~Wang:
{\em Computers and Mathematics with Applications \/}
{\bf 58}, 104-118 (2009),
General identities on Bell polynomials.

\bibitem{appendix} See Supplemental Material at [URL will be inserted by publisher] for numerical simulation algorithm for an arbitrary kernel and for issues of theoretical calculations.


	
	 
	


	


	
	
	
	
	
	
	
	
	
	
	
	
	
	
	
	
	
	
		
	



	










\bibitem{2019_PRE_Fronczak} A.~Fronczak, M.~Łepek, P.~Kukliński, P.~Fronczak:
{\em Phys. Rev. E\/}
{\bf 99}, 012104 (2019),
Coagulation with product kernel and arbitrary initial conditions: Exact kinetics within the Marcus-Lushnikov framework.

\bibitem{2004_Menon} G.~Menon, R.L.~Pego:
{\em Commun. Pure Appl. Math. \/}
{\bf 57}, 1197-1232 (2004),
Approach to self-similarity in Smoluchowski's coagulation equations.










	
	




		
\end{thebibliography}
\end{document}


\section*{Appendix}

Here, we discuss the computational details of the calculations of the presented equations and the algorithm of the numerical simulation of an arbitrary aggregation kernel. The code for the simulations and for the theoretical predictions used in this work is available at https://github.com/mlepek/aggregation. The instructions on how to run and modify the code are included in the repository.

\subsection{Theoretical calculations}

The largest problem in applying theoretical predictions is an explosion of the number of digits of Bell polynomials for $N>100$ as this number immediately exceeds the precision available in standard programming environments. Thus, the calculations have to be done with the help of arbitrary precision computation packages. We used the GNU MPFR library \cite{mpfr} with the extension MPFRC++ \cite{mpfrcpp} for the C++ language.

The second vital problem is the computational time. To avoid calculation of solutions of Diophantine equation (the definition of Bell polynomial), which is NP-hard problem, it is more efficient to compute Bell polynomials using recursive relation 
\begin{equation} \label{Bell_recursion}  
 B_{n,k} = \sum_{m=1}^{n-k+1} \binom{n-1}{m-1} g_{m} B_{n-m,k-1},
\end{equation}
 
\noindent where $B_{0,0} = 1$, $B_{n,0} = 0$ for $n\ge1$, and $B_{0,k} = 0$ for $k\ge1$. Using this recursive equation, the time for Bell polynomial calculations was significantly reduced. For the single-thread software running on the single PC machine, the computational time needed to calculate $\langle n_s \rangle$ for all values of $s$ was about ten seconds. We believe this time to be relatively short, but, possibly, it can be further reduced by software or hardware optimizations or by parallel computing.

\subsection{Implementing an arbitrary kernel in numerical simulation}

The crucial part of the numerical simulation of the aggregating system with binary aggregation acts is implementation of the reaction kernel, $K(i,j)$. The probability of merging of two clusters of sizes $i$ and $j$ is proportional to the number defined by the kernel. Thus, in each time--step of the simulation, one must choose two particles to be merged with probability proportional to $K$. Now, we will show how to do this for several examples of kernels, including the electrorheological kernel. 

Let us start with very basic example of the constant kernel. Suppose, that we store the particles in a vector $V1$ and we are about to choose two particles from this vector of available particles to merge them. In the constant kernel, the probability is $K=1$, which means that all of the particles have the same probability of being chosen for the coagulation act. In this case, we can simply randomly choose (with uniform distribution) two (different) particles from the vector and merge them. This will implement the constant kernel.

Let us now consider the multiplicative (or product) kernel, where $K=ij$. In this case, the probability of choosing a particle for the coagulation act is proportional to its size. The easiest (and fastest) way of implementing this kernel is to produce a vector $V2$ which stores the particles present in the system, but each particle is stored multiple times -- proportional to its size. For instance, if we have 3 particles in the system, say, A of size 1, B of size 2 and C of size 5, the vector will store: [A, B, B, C, C, C, C, C]. Now, we can randomly choose (with uniform distribution) two cells of this vector and these two cells will indicate which two particles we shall merge. This will implement the multiplicative kernel. 

It is important to note, that when you choose two cells from the vector, you shall check if they are related to \textit{different} particles (here: different letters) and discard your choice if they are from the same particle. In this step, it is vital that you choose both cells at once, check if they are from different particles, and discard both if they are not. The other scenario of choosing, when you choose one cell, and then you choose the second cell repeatedly until it is from the other particle, is depreciated as it changes the probabilities. If these two scenarios are compared, (slightly) different results are observed.

It was shown in literature that the additive kernel can be implemented by choosing randomly one cell from the vector $V1$ and one cell from the vector $V2$ \cite{additive_kernel, 2019_ROMP_Lepek}. These two cells indicate the two particles to be merged.

The next example of a kernel, we will discuss, is an arbitrary kernel which gives integer numbers but cannot be easily implemented as the three kernels mentioned above (or, at least, such an implementation is not known). As an example, we can take $K=1+i+j+ij$. The solution to this case is to create a vector of possible coagulation acts. Suppose, again, that we have 3 particles in our system, A of size 1, B of size 2 and C of size 5. The possible coagulation acts are: A+B, A+C, and B+C. For each pair of particles, the integer probability $K$ of this coagulation act can be calculated. For the pair A+B, the probability is $K=6$, for the pair A+C, the probability is $K=9$, etc. Therefore, we construct a vector containing possible pairs of particles where a particular pair occurs $K$--times in the vector. For this case, the vector will consist of: [A+B, A+B, A+B, A+B, A+B, A+B, A+C, A+C, A+C, A+C, A+C, A+C, A+C, A+C, A+C, B+C, B+C, B+C, B+C, B+C, B+C, B+C, B+C, B+C, B+C, B+C, B+C, B+C, B+C, B+C, B+C, B+C]. By choosing randomly one cell from this vector, we determine which two particles shall be merged.

But, what if the kernel expression gives real numbers instead of integers, e.g. the ER kernel, $K=\frac{1}{i}+\frac{1}{j}$? The method described in the previous paragraph is no longer usable. In this case, what can we do? The answer to this question is, that we must modify our vector to store probabilities of coagulation acts, instead of storing these coagulation acts themselves. Let us suppose, again, that we have 3 particles in our system, A of size 1, B of size 2 and C of size 5. As we are regarding the ER kernel, the probability of merging A and B is 1.5, the probability of merging A and C is 1.2, and the probability of merging B and C is 0.7. Now, we construct a vector which contains cumulative sums of the calculated probabilities. In the first cell we put the probability of pair A+B, in the second cell we put the sum of probabilities of A+B and A+C, and in the last cell we put the sum of probabilities of all pairs. Thus, our vector consists of three cells: [1.5, 1.5+1.2, 1.5+1.2+0.7], which, finally, is [1.5, 2.7, 3.4]. The last step we must perform is to randomly choose (with uniform distribution) a real number from the range from 0.0 to 3.4, i.e., the maximal (last) value in the vector. This random number will indicate one particular cell in the vector. Let us suppose we have randomly chosen the number of 1.9. We check, if it is lower or equal to the first cell (1.5). It is higher, thus, we go to the next cell. This time, the number of 1.9 is lower than the probability in the second cell of the vector (2.7), thus, it is the cell of interest. We know that in this second cell, we added the probability of the pair of A+C, therefore, we shall take the pair A+C as the coagulation act in this time--step of our simulation. This way, we have implemented the ER kernel. The method described in this paragraph is valid for any arbitrary kernel.